\documentclass[pra,aps,amsmath,showpacs,preprintnumbers]{revtex4}
\usepackage{graphicx}
\usepackage{bm}
\usepackage{tabularx}

\newcommand{\mE}{\mathcal{E}}
\newcommand{\mF}{\mathcal{F}}
\newcommand{\mT}{\mathcal{T}}
\newcommand{\pt}{\partial}

\begin{document}

\title{Dynamics of rapidly rotating Bose-Einstein condensates
       in a harmonic plus quartic trap}

\author{Jong-kwan Kim$^{1,3}$ and Alexander L. Fetter$^{1,2,3}$ }
\affiliation{$^{1}$Geballe Laboratory for Advanced Materials,
                   Stanford University,
                   Stanford, CA 94305-4045\\
             $^{2}$Department of Physics,
                   Stanford University,
                   Stanford, CA 94305-4060\\
             $^{3}$Department of Applied Physics,
                   Stanford University,
                   Stanford, CA 94305-4090}

\date{May 3, 2005}

\begin{abstract}
A two-dimensional rapidly rotating Bose-Einstein condensate
in a harmonic plus quartic trap
is expected to have unusual vortex states
that do not occur in a pure harmonic trap.
At a critical rotation speed $\Omega_h$,
a central hole appears in the condensate,
and at some faster rotation speed $\Omega_g$,
the system undergoes a transition
to a \emph{giant} vortex state with pure irrotational flow.
Using a time-dependent variational analysis,
we study the behavior of an annular condensate
with a single concentric ring of vortices.
The transition to a giant vortex state is investigated
by comparing the energy of the two equilibrium states
(the ring of vortices and the giant vortex)
and also by studying the dynamical stability
of small excitation modes of the ring of vortices.
\end{abstract}


\maketitle


\section{Introduction}
Since the creation of a single vortex in Bose-Einstein condensates
in harmonic traps \cite{Matthews1999a,Madison2000a},
experimental techniques have rapidly
created arrays of several hundred vortices
\cite{Abo-Shaeer2001a,Raman2001a,Haljan2001b,Engels2002a}.
In these experiments,
the condensate rotates at angular velocities $\Omega$
that typically approach the high-rotation limit $\omega_\perp$,
where the centrifugal effect from the rotation would cancel
the radial confinement of the trap
\cite{Raman2001a,Haljan2001b,Fetter2001a}.

For $\Omega\rightarrow\omega_\perp$
in a pure harmonic radial trap,
the combined effective radial trap vanishes
and the Thomas-Fermi (TF) radius of the condensate diverges.
Moreover, as Ho \cite{Ho2001a} pointed out,
the single-particle Hamiltonian in a rapidly rotating harmonic trap
is analogous to that of a charged particle in a uniform magnetic field,
and all particles should condense into the lowest Landau level (LLL).
This insight has stimulated several
experimental \cite{Schweikhard2004a,Coddington2004a}
and theoretical \cite{Baym2004a,Cooper2004a,Watanabe2004a} works.

An additional radial trap potential, stronger than harmonic,
opens new possibilities in the study
of rapidly rotating Bose-Einstein condensates,
because it can provide confinement even for $\Omega\ge\omega_\perp$.
A quartic potential is a simple choice
\cite{Fetter2001a,Lundh2002a}
that has recently been realized experimentally
with a blue-detuned laser directed along the axial direction
of an elongated condensate
\cite{Bretin2004a,Stock2004a}.
Not only does the combined potential overcome
the limit on the angular velocity $\Omega$,
but it also enriches
the physics of rapidly rotating condensates
with the possibility of the interesting new vortex configurations.

In such a combined potential,
no quantized vortices appear in the condensate
at sufficiently slow rotations $\Omega\ll\omega_\perp$.
With increasing $\Omega$ $\left(\lesssim\omega_\perp\right)$,
the condensate passes through a sequence of states
with many vortices that eventually form a large vortex lattice.
These states are essentially the same as those in a pure harmonic trap.
The new vortex phases begin to appear
as the condensate continues to expand radially for $\Omega>\omega_\perp$.
Specifically it develops a central hole
at a critical angular velocity $\Omega_h$
that depends on the interaction strength and the trap potential.
For larger values of $\Omega>\Omega_h$,
the condensate forms an annulus,
whose mean radius expands
and whose width continues to shrink with increasing $\Omega$.
Finally a ``giant'' vortex state with pure irrotational flow
is expected to appear at $\Omega_g$,
when all the singly quantized vortices disappear from the annulus.

Previous theoretical work has
either relied on numerical methods
\cite{Kasamatsu2002a,Kavoulakis2003a}
or dealt only with weak interactions and small anharmonicities
\cite{Jackson2004a,Jackson2004b}.
In contrast, we here use a time-dependent variational analysis
to study the behavior of an annular condensate
with a single ring of vortices in the annulus in the TF limit.
As $\Omega$ increases beyond $\Omega_h$,
the width of the annular condensate becomes increasingly narrow,
and above some value of angular speed $\Omega$
(smaller than $\Omega_g$)
the annular condensate can support only
a one-dimensional array of vortices arranged in a single ring
\cite{Kasamatsu2002a,Kim2004a,Fetter2005a}.
The transition to a giant vortex state is investigated
by comparing the energy of these two equilibrium states
(a state with a single ring of vortices
and a giant vortex state with pure irrotational flow)
and also by studying the dynamical stability of
small-amplitude normal mode excitations of the vortex array.
Section~\ref{sec:basic} summarizes the basic procedure
of the Lagrangian formalism
and the analytical study in the TF limit.
In Sec.~\ref{sec:variation},
we introduce our basic approximate condensate density profile,
which is simple, yet incorporates
the discrete character of the quantized vortices.
We analyze the equilibrium (Sec.~\ref{sec:variation})
and dynamical (Sec.~\ref{sec:Tkanchenko}) behavior of the condensate
using variational methods,
and the transition angular velocity $\Omega_g$
is evaluated numerically for various values
of the interaction parameter and the trap parameter.

\section{Basic Procedure}
\label{sec:basic}
A trapped Bose-Einstein condensate at zero temperature
can be described with a macroscopic order parameter
(the condensate wave function)
$\Psi\equiv\langle\hat{\Psi}\rangle$.
It satisfies the time-dependent, nonlinear Schr\"odinger equation
(Gross-Pitaevskii equation),
\begin{equation}
\label{eq:schrodinger_lab}
i\hbar\frac{\pt\Psi}{\pt t}
  = - \frac{\hbar^2\bm\nabla^2}{2M}\Psi
    + V_\mathrm{tr}\Psi
    + \frac{4\pi\hbar^2 a_s}{M} \left|\Psi\right|^2 \Psi,
\end{equation}
where $a_s$ is the positive $s$-wave scattering length,
characterizing the repulsive interaction
between the condensate particles.
With an additional quartic component,
the external two-dimensional trap potential $V_\mathrm{tr}$
can be written as
\begin{equation}
\label{eq:trap}
V_\mathrm{tr}
  = \frac{1}{2} M \omega_\perp^2 \left(r^2+\lambda\frac{r^4}{d_\perp^2}\right)
  = \frac{1}{2} \hbar\omega_\perp
                \left(\frac{r^2}{d_\perp^2}+\lambda\frac{r^4}{d_\perp^4}\right),
\end{equation}
where $d_\perp\equiv\sqrt{\hbar/M\omega_\perp}$
is the harmonic oscillator length,
$\mathbf{r}$ is the radial coordinate in two dimensions,
and $\lambda$ is a dimensionless parameter
that determines the relative strength of the quartic component.

If the external trap potential rotates at an angular velocity
$\bm\Omega=\Omega\hat{\mathbf{z}}$,
it is necessary to work in a co-rotating frame,
and the time-dependent Schr\"odinger equation becomes
\begin{equation}
\label{eq:schrodinger_rotating}
i\hbar\frac{\pt\Psi}{\pt t}
  = - \frac{\hbar^2\bm\nabla^2}{2M}\Psi
    + V_\mathrm{tr}\Psi
    + \frac{4\pi\hbar^2 a_s}{M} \left|\Psi\right|^2 \Psi
    - \bm\Omega\cdot\mathbf{L}\Psi,
\end{equation}
where
$\mathbf{L}
  = \mathbf{r}\times\mathbf{p}
  = -i\hbar\mathbf{r}\times\bm\nabla$ is
the angular momentum operator.

Frequently, instead of this Schr\"odinger equation,
it is preferable to use an equivalent Lagrangian formalism,
\begin{equation}
\label{eq:Lagrangian}
\mathcal{L}\left[\Psi\right]
  \equiv \mT\left[\Psi\right] - \mF\left[\Psi\right],
\end{equation}
where
\begin{equation}
\label{eq:Langrangian_time}
\mT\left[\Psi\right]
  \equiv \int dV ~
         \frac{i\hbar}{2}
         \left(\Psi^*\frac{\pt\Psi}{\pt t}-\frac{\pt\Psi^*}{\pt t}\Psi\right)
\end{equation}
is the time-dependent part of the Lagrangian functional and
\begin{equation}
\label{eq:Lagrangian_free}
\begin{split}
\mF\left[\Psi\right]
  &\equiv \mE'\left[\Psi\right] - \mu N \\
  &= \int dV
     \left[\frac{\hbar^2}{2M}\left|\bm{\nabla}\Psi\right|^2
           + V_\mathrm{tr} \left|\Psi\right|^2
           + \frac{2\pi\hbar^2 a_s}{M} \left|\Psi\right|^4
           - \Psi^* \bm\Omega\cdot\mathbf{L}\Psi
           - \mu \left|\Psi\right|^2\right],
\end{split}
\end{equation}
is the free-energy functional, incorporating the constraint
of fixed particle number $N=\int dV \left|\Psi\right|^2$
with the chemical potential $\mu$ as a Lagrange multiplier.

For simplicity, we consider an effectively two-dimensional system,
uniform in the $z$ direction over a length $Z$.
The condensate wave function can be chosen as $\Psi=\sqrt{N/Z}\psi$,
and the normalization condition becomes
\begin{equation}
\label{eq:norm_dimless}
1 = \int d^2\mathbf{r} \left|\psi\right|^2.
\end{equation}
In dimensionless form,
with $d_\perp$ and $\omega_\perp$ as scales for length and frequency,
the free-energy functional (per particle) becomes
\begin{equation}
\label{eq:energy_dimless}
\mF = \int d^2\mathbf{r}
      \left[\frac{1}{2}\left|\bm\nabla\psi\right|^2
            + \frac{1}{2}\left(r^2+\lambda r^4\right) \left|\psi\right|^2
            + \frac{g}{2} \left|\psi\right|^4
            + i\psi^*\bm\Omega\cdot\mathbf{r}\times\bm\nabla\psi
            - \mu\left|\psi\right|^2 \right],
\end{equation}
where $g=4\pi N a_s/Z$ is a dimensionless coupling constant,
characterizing the interaction strength between the condensate particles.
The physics of the condensate is now determined by two dimensionless constants:
the quartic trap strength $\lambda$ and the interparticle strength $g$.

Using the replacement $\psi=\sqrt{n} e^{iS}$,
we can write the first term on the right hand side
in Eq.~(\ref{eq:energy_dimless}) as
$\left|\bm\nabla\psi\right|^2=\left|\bm\nabla\sqrt{n}\right|^2
    +\left|\bm\nabla S\right|^2 \left|\psi\right|^2$.
In our units, the dimensionless particle velocity field
is given by $\mathbf{v}=\bm\nabla S$
and, in the Thomas-Fermi approximation,
the curvature of the density $\bm\nabla\sqrt{n}$ is neglected.
In this limit, variation of the free energy $\mF$
with respect to $\left|\psi\right|^2$
yields the familiar TF density
\begin{equation}
\label{eq:TF_original}
g \left|\psi\right|^2
  = \mu + \bm\Omega\cdot\mathbf{r}\times\mathbf{v}
    - \frac{1}{2}\left(v^2+r^2+\lambda r^4\right).
\end{equation}
Given the velocity $\mathbf{v}$,
we can determine $\mu$ with the normalization condition,
Eq.~(\ref{eq:norm_dimless}).

\subsection{Uniform vortex lattice with central hole}
In the limit of a dense uniform vortex lattice
with dimensionless areal density $\Omega/\pi$ \cite{Tilley1990a},
the total velocity field arising from
each quantized vortex in the annular region $R_<\le r\le R_>$
can be approximated by an integral, yielding
\begin{equation}
\mathbf{v}_\mathrm{a}\left(\mathbf{r}\right)
  = \Omega \left(r-\frac{R_<^2}{r}\right) \hat{\bm\phi}.
\end{equation}
The first term is the anticipated solid-body rotation,
but the second term reflects the presence of the central hole in the annulus.
To better mimic solid-body flow
that minimizes the free energy in the annulus,
it is necessary to add the irrotational flow
$\mathbf{v}_\mathrm{irr}\left(\mathbf{r}\right)
  = \left(\Omega R_<^2/r\right)\hat{\bm\phi}$,
arising from the phantom vortices in the hole.
Then the total superfluid flow in the annular region becomes
the sum of the contribution $\mathbf{v}_\mathrm{a}$
from vortices in the annulus
and the irrotational flow
$\mathbf{v}_\mathrm{irr}$ from the phantom vortices,
$\mathbf{v}=\mathbf{v}_\mathrm{a}+\mathbf{v}_\mathrm{irr}
           =\bm\Omega\times\mathbf{r}=\Omega r\hat{\bm\phi}$,
which is the expected solid-body flow.

Substituting this velocity field,
the TF density in Eq.~(\ref{eq:TF_original}) is simplified to
\begin{equation}
g \left|\psi\right|^2
  = \mu + \frac{1}{2}\left[\left(\Omega^2-1\right)r^2-\lambda r^4\right].
\end{equation}
This density profile predicts
the appearance of a central hole in the condensate
at angular speed $\Omega_h$ that is given by \cite{Fetter2005a}
\begin{equation}
\Omega_h^2
  = 1 + 2\sqrt{\lambda}\left(\frac{3\sqrt{\lambda}g}{2\pi}\right)^{1/3}.
\end{equation}
For $\Omega>\Omega_h$,
the squared TF inner and outer radii of the annulus
satisfy the simple relations
\begin{equation}
\label{eq:radii_relation}
R_>^2+R_<^2=\frac{\Omega^2-1}{\lambda}, \qquad
R_>^2-R_<^2=\frac{2}{\sqrt{\lambda}}
            \left(\frac{3\sqrt{\lambda}g}{2\pi}\right)^{1/3}.
\end{equation}

\subsection{Irrotational flow in annular region (``giant vortex'')}
The superfluid velocity field of pure irrotational flow
is simply $\mathbf{v}=\left(\nu/r\right) \hat{\bm\phi}$,
where $\nu$ is the quantum number
of the central circulation ($\nu$ is an integer.)
At sufficiently fast angular velocity,
$\nu$ will be large,
and we can treat $\nu$ as a continuous variable.
The TF density for a giant vortex is then given by
\begin{equation}
\label{eq:TF_irr}
g \left|\psi\right|^2
  = \tilde\mu - U\left(r^2\right),
\end{equation}
where $\tilde\mu\equiv \mu+\Omega\nu$ and
\begin{equation}
U\left(x\right) = \frac{1}{2}\left(\frac{\nu^2}{x}+x+\lambda x^2\right)
\end{equation}
can be considered an effective potential
that combines the centrifugal barrier
and the original trap potential (here, $x=r^2$).

\section{Variational Analysis}
\label{sec:variation}
As explained in the Introduction,
the width of the condensate becomes narrower
with increasing angular velocity,
and at some point,
the annular condensate can support
only a one-dimensional array of vortices arranged in a single ring.
We use a variational analysis
to study a condensate with irrotational flow
around a central hole (the giant vortex state)
and a condensate with a single ring of $N_a$ singly quantized vortices
in the annulus (the $N_a\neq0$ state).

\subsection{Simple parabolic density approximation}
Although the TF density profile of a giant vortex state
in Eq.~(\ref{eq:TF_irr})
is simple and manageable,
when we include vortices in the annulus,
it becomes complicated and almost untractable
for variational analysis.
We need a simpler density approximation,
yet it should contain the same physics as the full system.

From the TF density profile of a giant vortex state
(\ref{eq:TF_irr}),
we know that the condensate is confined in an annular region
bounded by inner and outer radii,
and its density profile is a function of $r^2$.
As a simple model of the density profile for an annular condensate,
we use the following parabolic functional form.
\begin{equation}
\label{eq:density_para}
g n
  = g \left|\psi\right|^2
  = A (X_>-r^2)(r^2-X_<),
\end{equation}
where $X_<$ and $X_>$ are variational
inner and outer squared radii
of the condensate, respectively,
to be determined such that they minimize
the free energy of the condensate.
A positive constant $A>0$ is
also introduced to ensure the normalization condition.

\begin{figure}
\centering
\includegraphics[height=5in,width=6.25in]{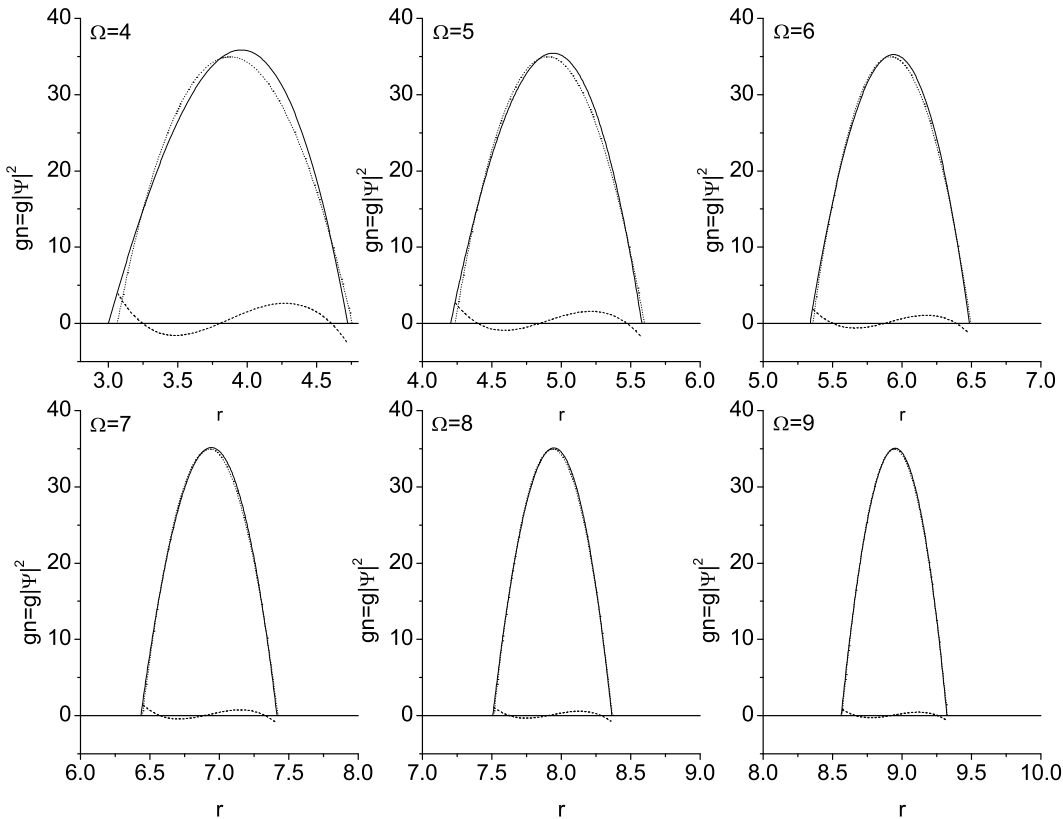}
\caption{The simple parabolic density (\ref{eq:density_para})
         and the TF density (\ref{eq:TF_irr}) profiles
         for $\lambda=1/2$ and $g=1000$,
         at $\Omega=\left(4,5,6,7,8,9\right)$.
         The parabolic profiles are shown in solid lines,
         the TF profile in dotted lines,
         and, for comparison, the difference between the two is dashed.
         In all graphs,
         the horizontal axes (radius $r$) are drawn in the same scale,
         and we can easily see
         that the mean condensate radius becomes bigger
         and the width of the condensate becomes narrower
         with increasing $\Omega$.
        }
\label{fig:density_profile}
\end{figure}

The comparison to the TF density profile
of a giant vortex state is shown
in Fig.~\ref{fig:density_profile}
for $\lambda=1/2$ and $g=1000$, from $\Omega=4$ to $\Omega=9$.
The optimized density profile from Eq.~(\ref{eq:density_para})
is plotted in solid lines
and compared to the TF profile (dotted).
For easy comparison,
the difference of two profiles (the parabolic density and the TF density)
is also shown in dashed lines.
Even at low $\Omega$ values $\left(\text{for example~}\Omega=4\right)$,
this simple approximation shows good agreement
-- when the difference is compared to the maximum condensate density,
it is $\lesssim$~5\%.
The approximation gets better
as the external rotation speed $\Omega$ becomes faster,
and at $\Omega=6$, the two density profiles are almost the same.
Generally, the parabolic density profile has
smaller inner and outer radii,
but a slightly higher peak.

We start from this trial parabolic density profile,
along with an appropriate velocity field,
and calculate the free-energy functional $\mF$.
Later we repeat the procedure with a different improved model.
As in \cite{Fetter2005a},
the irrotational flow in the annulus
can be approximated by the central circulation $\nu$.
The total velocity field given by the phase $S$ is
\begin{equation}
\mathbf{v}
  = \bm\nabla S
  = \frac{\nu}{r} \hat{\bm\phi} + \sum_k \mathbf{v}_k,
\end{equation}
where $\mathbf{v}_k$ is the contribution
from the $k$th vortex located at $\mathbf{r}_k=\left(r_k,\phi_k\right)$
as one of a single ring of vortices in the annular region,
\begin{equation}
\mathbf{v}_k
  = \hat{\mathbf{z}} \times
    \frac{\mathbf{r}-\mathbf{r}_k}{\left|\mathbf{r}-\mathbf{r}_k\right|^2}.
\end{equation}

In terms of the trial wave function $\psi=\sqrt{n} e^{iS}$,
the free-energy functional $\mF$ is now characterized
by a set of parameters,
$\mF=\mF\left(\mu,\nu,X_<,X_>,A,\{\mathbf{r}_k\}\right)$.
For a given $\Omega$, we choose a value of $N_a$
(since the number of vortices $N_a$ is not a continuous variable)
and we then minimize $\mF$ with respect to all parameters other than $\mu$,
\begin{equation}
\label{eq:minimize}
0 = \frac{\pt\mF}{\pt\nu}
  = \frac{\pt\mF}{\pt X_<}
  = \frac{\pt\mF}{\pt X_>}
  = \frac{\pt\mF}{\pt A}
  = \frac{\pt\mF}{\pt r_k}
  = \frac{\pt\mF}{\pt \phi_k},
\end{equation}
along with the normalization condition
that determines $\mu$ with Eq.~(\ref{eq:minimize}),
\begin{equation}
1 = \frac{\pi}{6} \frac{A}{g} \left(X_>-X_<\right)^3.
\end{equation}
Once we minimize $\mF$ and have all the values of the other parameters
for given $\left(\Omega,N_a\right)$,
then we can compare the energy $\mE'=\mF+\mu$
among different $N_a$ vortex states
to determine the lowest energy state (ground state) for a given $\Omega$.

Table \ref{tbl:simple} shows (meta)stable states
for some typical values of $\Omega$
at $\lambda=1/2$ and $g=1000$.
Here we only consider $N_a\neq 0$ states,
omitting the giant vortex state from the table.
$\bar{R}\equiv\left(\sqrt{X_>}+\sqrt{X_<}\right)/2$
is the mean radius of the condensate,
$R_r$ is the variational radius $\left(r_k\right)$
of a symmetrically spaced ring of vortices in the annular region,
$b$ is the intervortex spacing $b\equiv2\pi R_r/N_a$,
and $d\equiv\sqrt{X_>}-\sqrt{X_<}$ is the width of the condensate.
The conventional (usual) healing length
$\xi_0\equiv 1/\sqrt{2gn_0}$
is determined from the maximum density of the condensate
$n_0=\left(A/4g\right) \left(X_>-X_<\right)^2$.
The healing length $\xi_0$ in Table~\ref{tbl:simple}
is nearly constant,
since the maximum density $n_0$ remains almost the same
as seen in Fig.~\ref{fig:density_profile}.
For comparison, the result from a numerical analysis
from Ref.~\cite{Fetter2005a} is shown in Table~\ref{tbl:Jackson}.

\begin{table}
\begin{ruledtabular}
\begin{tabular}{ccccccc}
$\Omega$ & $N_a$ & $\bar{R}$ & $R_r$ & $b$   & $d$  & $\xi_0$ \\
\hline
5        & 45    & 4.82      & 4.86  & 0.678 & 1.88 & 0.138 \\
6        & 57    & 5.87      & 5.89  & 0.650 & 1.51 & 0.138 \\
7        & 71    & 6.90      & 6.92  & 0.612 & 1.29 & 0.139 \\
\end{tabular}
\end{ruledtabular}
\caption{Result of a simple parabolic approximation
         for $\lambda=1/2$ and $g=1000$:
         Here $N_a$ is the optimized vortex number
           in a ring for given $\Omega$.
         The mean radius is
           $\bar{R}\equiv\left(\sqrt{X_>}+\sqrt{X_<}\right)/2$,
         $R_r$ is the variational radius of a ring of vortices,
         $b\equiv2\pi R_r/N_a$ is the intervortex spacing,
         $d\equiv\sqrt{X_>}-\sqrt{X_<}$ is the width of the condensate,
         and $\xi_0\equiv 1/\sqrt{2gn_0}=\sqrt{2/A} \left(X_>-X_<\right)^{-1}$
           is the healing length.
        }
\label{tbl:simple}
\end{table}

\begin{table}
\begin{ruledtabular}
\begin{tabular}{ccccccc}
$\Omega$ & $N_a$ & $R_\mathrm{meas}$ & $R_\mathrm{TF}$ & $b_\mathrm{meas}$ &
           $d_\mathrm{meas}$   & $d_\mathrm{TF}$     \\
\hline
5 & 37 & 4.83 & 4.79 & 0.821 & 2.32 & 2.06 \\
6 & 44 & 5.80 & 5.86 & 0.828 & 1.98 & 1.68 \\
7 & 51 & 6.85 & 6.89 & 0.844 & 1.76 & 1.43 \\
\end{tabular}
\end{ruledtabular}
\caption{Result of a numerical analysis
         for $\lambda=1/2$ and $g=1000$ \cite{Fetter2005a}:
         $N_a$ here is determined
         by solving the time-dependent GP equation.
         $R_\mathrm{meas}$, $b_\mathrm{meas}$ and $d_\mathrm{meas}$
         are the mean radius, the intervortex spacing
         and the width of the annulus
         from the numerical calculation, respectively.
         Note that the solution of the full time-dependent GP equation
         does not have well-defined inner/outer radii
         where the condensate density vanishes.
         Thus, the determination of $d_\mathrm{meas}$ depends on
         the numerical technique used to identify the radii,
         and only the qualitative comparison is meaningful.
         For comparison,
         the TF estimates from Eq.~(\ref{eq:radii_relation})
         are also shown:
         $R_\mathrm{TF}=\left(R_>+R_<\right)/2$
         and $d_\mathrm{TF}=R_>-R_<$.
        }
\label{tbl:Jackson}
\end{table}

As seen in Table~\ref{tbl:simple},
our variational study with the simple parabolic
approximation in Eq.~(\ref{eq:density_para}) predicts
that the intervortex spacing $b$ decreases with increasing $\Omega$,
although somewhat slower than the dependence $\Omega^{-1/2}$
associated with a uniform vorticity.
In contrast,
the numerical study of the full GP equation \cite{Fetter2005a} found
that the intervortex spacing $b$ \emph{increases} slowly
with increasing external rotation $\Omega$.
Evidently, the simple parabolic approximation
tends to overestimate the number of vortices in equilibrium.
This qualitative feature must be corrected
if we expect a realistic result from our model.

\subsection{Modified parabolic density approximation}
We can improve the simple parabolic model
by including the contribution
of the nonuniform density near the core of each vortex
to the total free energy.
This effect makes the addition of a vortex
in the annular region more costly
and thus tends to reduce the number of vortices in equilibrium,
increasing the intervortex spacing $b$.
This core contribution becomes more significant
as the radial trap rotates faster
and the number of vortices increases.

As a model for the vortex core structure,
we choose a simple linear model
\cite{Fischer2003a,Kavoulakis2003a,Baym2004a}.
\begin{equation}
\label{eq:density_mod}
\psi = \sqrt{n} e^{iS} F,
\end{equation}
\begin{equation}
F \equiv \prod_k f_k,
\end{equation}
where $n$ is the parabolic density in Eq.~(\ref{eq:density_para}) and
\begin{equation}
\label{eq:mod_func}
f_k = \begin{cases}
      1,    \quad
            & \text{if $\left|\mathbf{r}-\mathbf{r}_k\right|\ge\xi$}\\
      \left|\mathbf{r}-\mathbf{r}_k\right|/\xi, \quad
            & \text{if $\left|\mathbf{r}-\mathbf{r}_k\right|<\xi$}
      \end{cases}
\end{equation}
is a modified condensate density
due to the presence of the vortex.
Here we assume that the cores of the vortices do not overlap.
We use the symbol $F$ for notational convenience
to denote the product.
Note that $\xi$ is now a variational parameter
that determines the size of the vortex core;
it should be distinguished from the previous healing length
$\xi_0=1/\sqrt{2gn_0}$
that is determined directly
from the maximum condensate density $n_0$.

With this modified density profile,
the new normalization condition becomes,
up to corrections of order $\xi^2$,
\begin{equation}
1 \approx \frac{\pi}{6} \frac{A}{g} \left(X_>-X_<\right)^3
          - \frac{\pi\xi^2}{2}\sum_k n\left(r_k^2\right),
\end{equation}
where $\xi^2$ is assumed small (measured in units of $d_\perp^2$).
We can recalculate the free-energy functional $\mF$,
using
\begin{equation}
\bm\nabla\psi
  = \left(i\bm\nabla S F\sqrt{n}
    + \bm\nabla F\sqrt{n} + F\bm\nabla\sqrt{n}\right) e^{iS},
\end{equation}
\begin{equation}
\label{eq:grad_psi}
\left|\bm\nabla\psi\right|^2
  = \left|\bm\nabla S\right|^2 F^2 n + \left|\bm\nabla F\right|^2 n
    + F^2 \left|\bm\nabla\sqrt{n}\right|^2
    + 2 F \sqrt{n} \bm\nabla F\cdot\bm\nabla\sqrt{n}.
\end{equation}

If we again neglect the terms involving $\bm\nabla\sqrt{n}$,
we find
\begin{equation}
\begin{split}
\mF
  &= \int d^2\mathbf{r}
          \left[\frac{1}{2}\left(v^2+r^2+\lambda r^4\right) n
                - \left(\Omega r v_\phi+\mu\right) n
                + \frac{g n^2}{2} \right] \\
  &\phantom{=~~}
     +\sum_k \int d^2\mathbf{r}
                  \left[\frac{1}{2}\left(v^2+r^2+\lambda r^4\right)
                        - \left(\Omega r v_\phi+\mu\right)\right]
                  n \left(\left|f_k\right|^2-1\right) \\
  &\phantom{=~~}
     +\sum_k \int d^2\mathbf{r}
                  \left[\frac{g n^2}{2} \left(\left|f_k\right|^4-1\right)\right] \\
  &\phantom{=~~}
     +\frac{1}{2} \int d^2\mathbf{r}
                  \left(\left|\bm\nabla F\right|^2 n \right).
\end{split}
\end{equation}
The first term has the same functional form
as the free energy of the simple parabolic density approximation
in the previous subsection;
in contrast, the last three terms come
from the core modification in the density.
The contribution to the free-energy integral from
the last term in Eq.~(\ref{eq:grad_psi}) approximately averages to zero,
since each $\bm\nabla f_k$ is cylindrically symmetric
with respect to the vortex position $\mathbf{r}_k$
and also because
$\bm\nabla\sqrt{n}\approx 0$
near the vortex position.

As in the simple parabolic density approximation,
we can again find the lowest energy (ground) state.
The only difference is
that we have one more variational parameter $\xi$,
determined by the condition $0=\pt\mF/\pt\xi$.

\begin{figure}
\centering
\includegraphics[height=6in,width=5in]{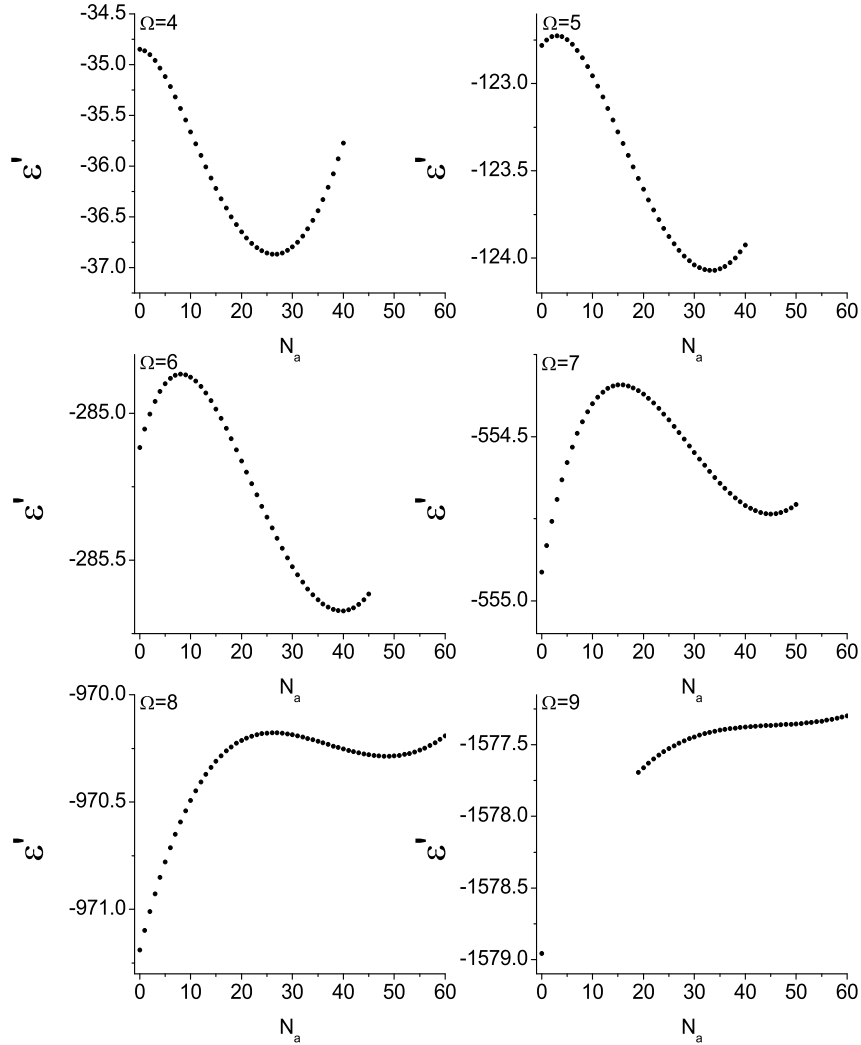}
\caption{Energy (per particle) $\mE'$ of various $N_a$ states
         for $\lambda=1/2$ and $g=1000$,
         at $\Omega=(4,5,6,7,8,9)$.
         Note that, for $\Omega=9$,
         we cannot find consistent solutions
         for $1\le N_a\le 18$.
        }
\label{fig:EvsNa}
\end{figure}

Figure \ref{fig:EvsNa} shows the energy $\mE'$
in the rotating frame
at several distinct values of $\Omega=(4,5,6,7,8,9)$
of various $N_a$ vortex states
that minimize $\mF$ with respect to the other parameters.
At $\Omega=4$, for example,
a giant vortex state $\left(N_a=0\right)$ is clearly unstable
and $N_a=26$ state is the lowest state
among the states with a single ring of vortices.
A faster external rotation ($\Omega=5 \text{ or } 6$)
makes a giant vortex state a local minimum
and therefore metastable
although the true minimum has $N_a\neq0$.
In the $\Omega=7$ figure,
a giant vortex state is the ground (lower-energy) state
and the $N_a=45$ state is now
only a local minimum and therefore metastable.
Finally in the $\Omega=9$ figure,
only a giant vortex state is stable,
for the local minimum at finite $N_a$ no longer appears.

The ground-state (or metastable-state) energy values
that we find through this variational procedure for $4\le\Omega\le7$
with the modified parabolic density approximation
agree well with the numerical result from the full GP equation
\cite{Fetter2005a}.
We find that in general our model has lower energy (more negative)
with a small discrepancy $\left(\lesssim 5\%\right)$,
but it is probably expected
because our TF model ignores
the (positive) energy contribution
of the condensate density curvature (\ref{eq:grad_psi}).
In this sense, it may be more meaningful
to compare the energy differences
between the irrotational giant vortex state $\left(N_a=0\right)$
and the annular $N_a\neq0$ state
(or the vortex lattice state in \cite{Fetter2005a}),
through which the TF effect effects should somewhat cancel out.
In our model,
$\triangle\mE'\equiv\mE'_{N_a\neq0} - \mE'_{N_a=0}$
increases at an almost constant rate of
$\sim$ 0.73 per $\Omega$,
starting from a negative value;
in contrast, although not as constant as in our model,
$\triangle\mE'$ in the numerical result \cite{Fetter2005a}
increases at a slower rate of $\sim$ 0.64 per $\Omega$.

Table \ref{tbl:modified} shows results
from the modified parabolic density approximation
for some typical values of $\Omega$.
In this modified model,
the energy of one vortex includes the core contribution.
As we have anticipated,
the core modification in the density reduces
the number of vortices in equilibrium
relative to the simple parabolic approximation.
In fact, the values are somewhat underestimated by 4$\sim$6
when compared to the full numerical solution of \cite{Fetter2005a},
and the intervortex spacing $b$ now increases
with increasing $\Omega$,
in agreement with the numerical result \cite{Fetter2005a}.
The variational core size $\xi$ is also shown in the table.
While the conventional healing length $\xi_0$
is basically determined by the balance
between the kinetic energy term
and the interatomic interaction term in the free energy,
the variational core size $\xi$ is determined by
the velocity field (the phase contribution to the kinetic energy),
the confining trap geometry and the interaction terms
within the TF limit.
When compared to the conventional healing length $\xi_0$,
we find that the ratio $\xi/\xi_0 \approx \left(1.73\sim 1.91\right)$.
For reference,
the experiments with a small quartic trap potential
$\left(\lambda\sim 10^{-3}\right)$
and no central hole
\cite{Schweikhard2004a,Coddington2004a}
found that $\xi/\xi_0\approx 1.94$.

\begin{table}
\begin{ruledtabular}
\begin{tabular}{cccccccc}
$\Omega$ & $N_a$ & $\bar{R}$ & $R_r$ & $b$   & $d$  & $\xi$ & $\xi_0$ \\
\hline
5        & 33    & 4.82      & 4.80  & 0.915 & 1.83 & 0.248 & 0.130\\
6        & 40    & 5.87      & 5.85  & 0.918 & 1.53 & 0.236 & 0.131\\
7        & 45    & 6.90      & 6.87  & 0.959 & 1.31 & 0.226 & 0.131\\
\end{tabular}
\end{ruledtabular}
\caption{Result of the modified parabolic density approximation
         for $\lambda=1/2$ and $g=1000$:
         $\xi$ is the variational vortex core size
         and the other parameters are defined as in Table~\ref{tbl:simple}.
         The conventional healing length $\xi_0$
         is slightly different from Table~\ref{tbl:simple},
         because of the somewhat different parameters.
         While the optimized vortex number $N_a$
         is overestimated in the simple parabolic model
         (Table~\ref{tbl:simple})
         compared to the result of the numerical analysis
         (Table~\ref{tbl:Jackson}),
         the modified parabolic model instead
         underestimates $N_a$ by 4$\sim$6.
         Importantly,
         the spacing between the adjacent vortices $b$
         increases slowly with increasing $\Omega$,
         in qualitative agreement with the numerical finding
         from \cite{Fetter2005a}.}
\label{tbl:modified}
\end{table}

\begin{figure}
\centering
\includegraphics[height=5in,width=6.25in]{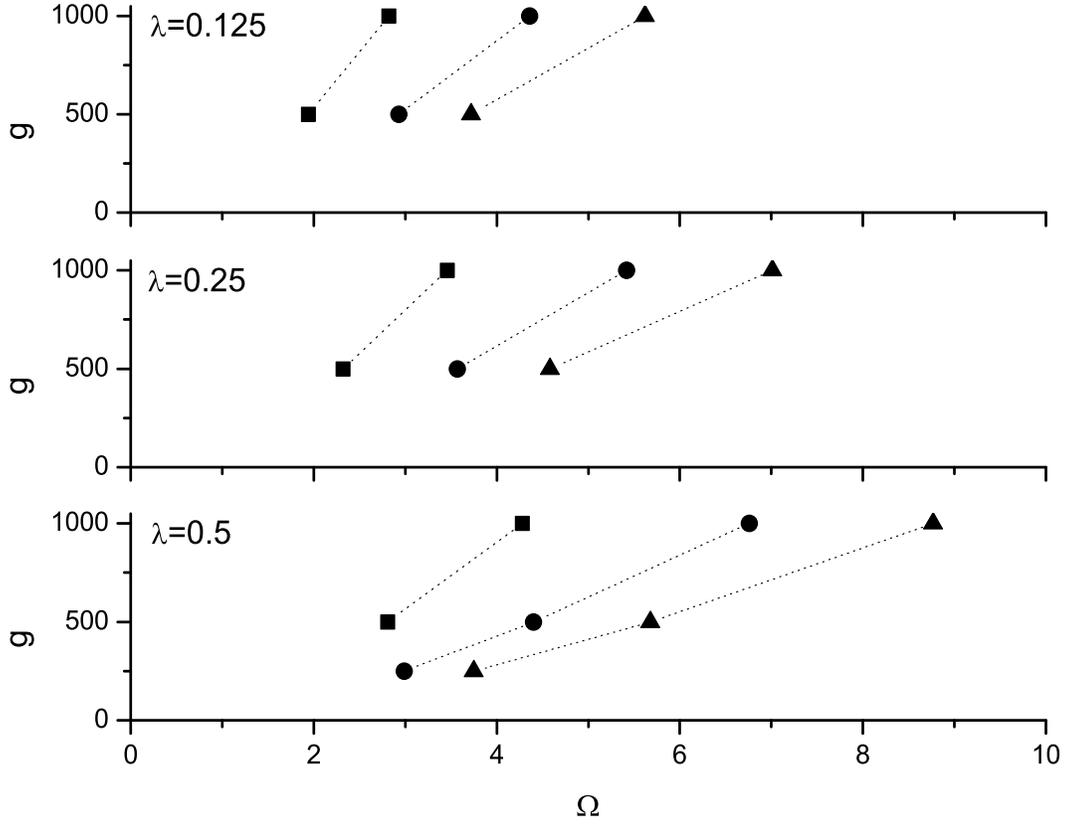}
\caption{Characteristic external rotation frequencies
         for $\lambda=(1/8,1/4,1/2)$ and $g=(250,500,1000)$.
         For $\Omega$ below the solid squares
         only states with $N_a\neq0$ are stable.
         Between solid squares and solid circles,
         giant vortex states are metastable
         and $N_a\neq0$ states are ground states.
         Between solid circles and solid triangles,
         the ground state is now the giant vortex state
         and $N_a\neq0$ states are only metastable.
         Above the solid triangles,
         only giant vortex states are stable.
        }
\label{fig:window}
\end{figure}

Figure \ref{fig:window} shows characteristic external rotation frequencies
for three anharmonic parameters $\lambda=(1/8,1/4,1/2)$
and a given interaction parameter $g=(250,500,1000)$.
Solid squares denote the external rotation frequencies
below which only states with a finite number
of vortices are stable (ground states)
and giant vortex states are not even metastable.
Between solid squares and circles,
finite-$N_a$ vortex states are ground states
and giant vortex states are metastable.
As the external rotation becomes faster,
at rotation frequencies given by solid circles,
a giant vortex state has the same energy
as the optimum finite-$N_a$ vortex state.
A giant vortex state becomes the ground state
for an external rotation frequency
faster than that denoted by a solid circle.
Finally at $\Omega$ faster than the solid triangle,
only the giant vortex state is stable
and presumably the ground state (at least within this model).

\begin{figure}
\centering
\includegraphics[height=6in,width=5in]{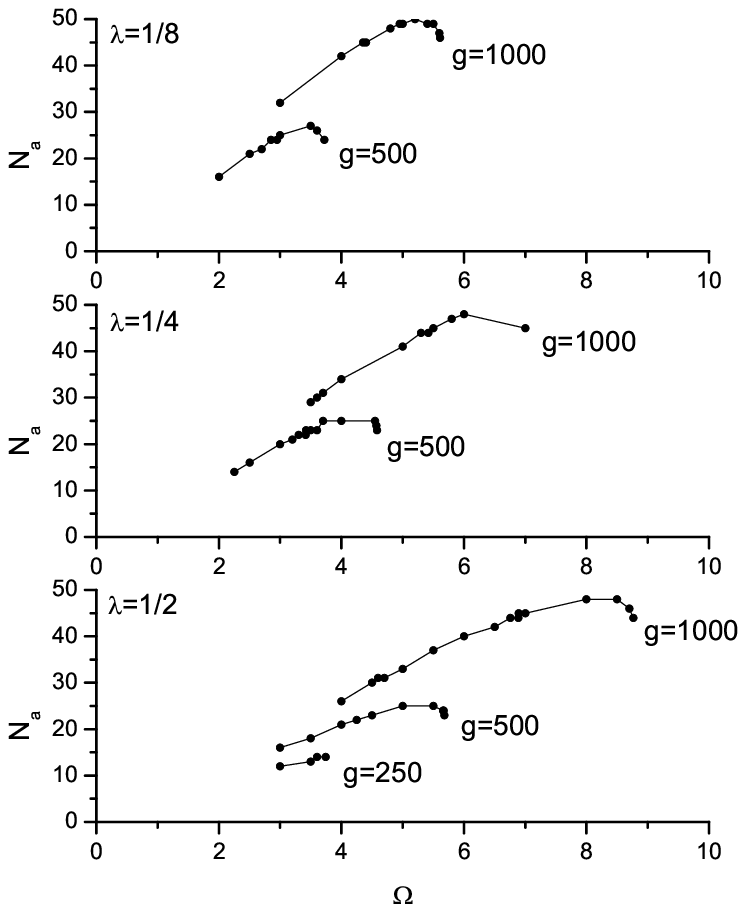}
\caption{Equilibrium number of vortices
         in a single ring in the annular region
         of a stable or metastable condensate
         as a function of the external rotation.
        }
\label{fig:NoVtx}
\end{figure}

Figure \ref{fig:NoVtx} shows the number of vortices
in the optimum symmetric ring in the annular region
of a stable or metastable condensate
as a function of the external rotation $\Omega$.
Interestingly, in most cases
above some relatively large value of $\Omega$
(although states with finite $N_a$ are no longer ground states),
the number of vortices in equilibrium indeed decreases
with the increasing external rotation
until the $N_a$ states become unstable.

\section{Tkachenko-like oscillation modes and the associated stability}
\label{sec:Tkanchenko}
By assumption, our equilibrium one-dimensional arrays of vortices
are symmetrically spaced around a circle.
From one of these equilibrium ground states with $N_a\neq0$,
we can consider the small-amplitude normal mode excitations
of the vortex array
by considering small variations about the equilibrium positions
of the vortices in the annulus.

The time-dependent part $\mT$
of the Lagrangian $\left(\mathcal{L}=\mT-\mF\right)$
can be calculated from
\begin{equation}
\begin{split}
\mT &\equiv \int d^2\mathbf{r}~
            \frac{i}{2}
            \left(\psi^*\frac{\pt\psi}{\pt t}
                  -\frac{\pt\psi^*}{\pt t}\psi\right)\\
    &= \sum_k \pi\dot\phi_k
              \left[-\frac{1}{6}\frac{A}{g}
                                \left(r_k^2-X_<\right)^2
                                \left(3X_>-2r_k^2-X_<\right)
                    -\frac{\xi^2}{2}\frac{A}{g}
                                    r_k^2 \left(X_>+X_<-2r_k^2\right)\right]\\
    &{\phantom{=~~}}
       + \sum_{j\neq k} \pi\dot\phi_k
                    \left[
                    \frac{\xi^2}{2} n\left(r_k^2\right)
                    \frac{r_j^2-r_j r_k\cos\phi_{jk}}
                         {r_j^2+r_k^2-2r_j r_k \cos\phi_{jk}}\right].
\end{split}
\end{equation}
As in our previous study of the dynamics of a single ring of vortices
in trapped Bose-Einstein condensates \cite{Kim2004a},
we introduce small variations $\alpha_k$ and $\beta_k$
for radial and angular coordinates of the $k$th vortex.
\begin{equation}
r_k = r_r + \alpha_k, \qquad
\phi_k = \phi_k^0 + \beta_k,
\end{equation}
where $\left(r_r, \phi_k^0\right)$
is the equilibrium position of the $k$th vortex
and $\phi_k^0=2\pi k/N_a$.

The Euler-Lagrange equations of motion
in terms of these small variations are
\begin{equation}
\label{eq:Euler_1}
\sum_k\left(\frac{\pt^2\mT}{\pt\dot\phi_k \pt r_q}\right) \dot\beta_k
  = \sum_k\left(\frac{\pt^2\mF}{\pt r_k \pt r_q}\right) \alpha_k
    + \sum_k\left(\frac{\pt^2\mF}{\pt \phi_k \pt r_q}\right) \beta_k,
\end{equation}
and
\begin{equation}
\label{eq:Euler_2}
\sum_k\left(\frac{\pt^2\mT}{\pt r_k \pt\dot\phi_q}\right) \dot\alpha_k
+ \sum_k\left(\frac{\pt^2\mT}{\pt\phi_k\pt\dot\phi_q}
              - \frac{\pt^2\mT}{\pt\dot\phi_k\pt\phi_q}\right) \dot\beta_k
  = -\sum_k\left(\frac{\pt^2\mF}{\pt r_k \pt \phi_q}\right) \alpha_k
    -\sum_k\left(\frac{\pt^2\mF}{\pt \phi_k \pt \phi_q}\right) \beta_k,
\end{equation}
where $1\le q\le N_a$.
In equilibrium, the $N_a$ vortices are symmetrically placed,
and the condensate has $N_a$-fold rotational symmetry.
As a result, a Fourier decomposition of $\bm\alpha$ and $\bm\beta$
\begin{equation}
\begin{pmatrix} \alpha_k \\ \beta_k \end{pmatrix}
  = \sum_{s=0}^{N_a-1} \begin{pmatrix}
                         \tilde\alpha_s \\ \tilde\beta_s
                       \end{pmatrix}
                       e^{i 2\pi k s /N_a},
\end{equation}
will decouple the original problem
into $N_a$ sets of $2\times 2$ matrix problems,
\begin{equation}
\frac{d}{dt} \mathbf{P}_s = M_s \mathbf{P}_s,
\end{equation}
where $\mathbf{P}_s=\left(\tilde\alpha_s,\tilde\beta_s\right)$.
The Fourier index $s$ is analogous to the wave number
in a linear array.
The matrix elements of $M_s$ are determined
from Eqs.~(\ref{eq:Euler_1}) and (\ref{eq:Euler_2}).

By solving these matrix problems,
we can study the dynamical stability of the vortex system.
For a given equilibrium state,
we determine the excitation spectrum
as a function of the external rotation frequency $\Omega$.
If a particular excitation frequency is real,
the corresponding small amplitudes $\left(\bm\alpha,\bm\beta\right)$
oscillate around zero,
and the state is stable.
If any excitation frequency is complex with a positive imaginary part,
the corresponding mode will grow with time,
and the state is dynamically unstable.

We find that all frequencies of these oscillation modes
are real for all annular states with $N_a\neq0$ in equilibrium;
thus the single ring of vortices in our annular condensate
is dynamically stable under small oscillations
of the vortex positions.
Note that this conclusion requires a detailed dynamical analysis.
In other cases, for example
a single ring of vortices in a circular two-dimensional condensate,
the ring becomes dynamically unstable
for sufficiently large vortex number \cite{Kim2004a}.

Our approximate variational Lagrangian trial function
treats only the vortex positions as dynamical variables.
In particular, we do not allow overall density variations.
Thus this analysis differs from that in \cite{Cozzini2005a},
where linearized rotational hydrodynamics is used
to analyze the low-lying normal modes of annular condensates.
Comparison of these two approaches merits further study.

\section{Transition to a giant vortex state}
\label{sec:transition}
According to the above study,
the transition from an annular state with $N_a\neq0$
to a giant vortex state
is determined from the energetic consideration
of the equilibrium states.
It is not determined from the dynamical stability
of the small-amplitude normal mode excitations
of the vortex lattice.

In the equilibrium analysis,
we can identify two characteristic rotational frequencies
that can be related to the transition;
one is the frequency
at which the ground state changes
from the annular $N_a\neq0$ state to the giant vortex state
(denoted by solid circles in Fig.~\ref{fig:window}),
and the other is the frequency
above which the annular vortex state is unstable
(denoted by solid triangles in Fig.~\ref{fig:window}).
Between these two frequencies,
the $N_a\neq0$ state and the giant vortex state
are energetically stable or metastable,
and dynamically stable.
As Jackson and Kavoulakis \cite{Jackson2004b} pointed out,
the existence of multiple (meta)stable states indicates
possible hysteresis effect in the phase transition
as the rotational angular frequency varies.
But strictly,
in the presence of some disturbing process/noise,
small enough not to destroy the system
yet large enough to overcome the energy barrier
between the two states,
the condensate seeks the global energy minimum state.
Thus we choose to identify the transition frequency $\Omega_g$
to the giant vortex as that
where the change of ground states occurs
(solid circles in Fig.~\ref{fig:window}).

\section{Discussion}
\label{sec:discussion}
We have used a variational analysis to study
a rapidly rotating, two-dimensional annular Bose-Einstein condensate
in a harmonic plus quartic radial trap,
concentrating on the states with a single ring of vortices in the annulus.
We have used a modified parabolic density profile approximation
that incorporates the density change near the core of each vortex.
Several parameters, inner squared radius $X_<$, outer squared radius $X_>$
and vortex core size $\xi$ are varied
to minimize the free energy $\mF$, leading to the stable state.

The time-dependent variational analysis indicates that
a transition to a pure irrotational giant vortex state
is determined solely from energetic considerations,
and cannot be associated with the dynamical stability
of small oscillation modes of vortex positions in this model.
We identify the transition frequency $\Omega_g$
to the giant vortex state
by comparing the energy of two equilibrium states,
taking $\Omega_g$ as the frequency
where the ground (global energy minimum) state changes
from $N_a\neq0$ to $N_a=0$.

We assume the validity of the TF approximation,
ignoring the contribution of the condensate density curvature
to the free-energy functional.
Since the usual healing length $\xi_0$
and the variational vortex core size $\xi$
are small compared to the intervortex spacing $b$
and the width of the condensate annulus $d$
in all ranges of $\Omega$ used in this study,
it should be safe to use the TF approximation.
Our analysis also argues against the different scenario
for the transition to a giant vortex state at large $\Omega$,
where $d$ becomes too narrow $\left(d\sim\xi\right)$
to accomodate a vortex in the annulus
and the TF assumption breaks down \cite{Fetter2005a}.

We study systems of various values
of $g=(250,500,1000)$ and $\lambda=(1/8,1/4,1/2)$.
For the smallest value of $g=125$ at $\lambda=1/2$,
we cannot find even a metastable state
with a single ring of vortices;
only a giant vortex state is stable.
In contrast, the numerical study \cite{Kasamatsu2002a}
at the same value of $g$ and $\lambda$
indeed found a transition from vortex arrays to a giant vortex.
The difference may reflect
our reliance on the TF approximation,
which holds for larger $g$.
For $g=1000$ at $\lambda=1/8$,
our transition angular frequency
to the giant vortex state
is $\Omega_g=4.36$.
A calculation based on a uniform vortex lattice
\cite{Kavoulakis2003a}
for similar values of $g=360\pi$ and $\lambda=1/8$
finds a considerably smaller value of the transition frequency
$\Omega_g\approx 2.40$.
For $g=1000$ at $\lambda=1/2$,
we find the transition angular frequency $\Omega_g=6.76$
to the giant vortex state,
while the numerical simulation \cite{Fetter2005a} finds
no transition even at $\Omega=7$.
Presumably this discrepancy arises
from our approximate variational analysis,
whereas the full numerical study \cite{Fetter2005a}
relies on the full time-dependent GP equation.
It may also reflect the finite grid size used in \cite{Fetter2005a}.
In addition, the vortices enter from the outer edge
in the imaginary-time formalism,
and they might become trapped in a metastable state
with a ring of vortices,
instead of reaching a giant vortex with irrotational flow \cite{Jackson2005a}.

It would also be interesting to extend the present analysis
to the case of a negative harmonic term,
where the condensate is annular even for $\Omega=0$
\cite{Jackson2005a,Aftalion2004a}.

\begin{acknowledgments}
We are grateful to B.~Jackson for helpful comments
and for sharing his numerical data with us.
\end{acknowledgments}


\end{document}